%% file: paper.tex
\documentclass[runningheads]{llncs}
\usepackage{graphicx}
\usepackage{algorithm}
\usepackage[noend]{algpseudocode}
\usepackage{etoolbox}
\usepackage{amsmath}
\usepackage{amssymb}
\usepackage{subcaption}
\usepackage{booktabs}
\usepackage{multirow}
\usepackage{pgfplots}
\usepackage{pgfplotstable}
\usepgfplotslibrary{groupplots}

\pgfplotsset{
    discard if not/.style 2 args={
        x filter/.code={
            \edef\tempa{\thisrow{#1}}
            \edef\tempb{#2}
            \ifx\tempa\tempb
            \else
                
            \fi
        }
    }
}
\pgfplotsset{compat=newest}

\usepackage{tikz}
\usetikzlibrary{decorations.pathmorphing}
\usetikzlibrary{decorations.pathreplacing}
\usetikzlibrary{calc}
\usetikzlibrary{fit}
\usetikzlibrary{matrix}

\definecolor{nicegreen}{RGB}{39,174,96}
\definecolor{niceblue}{RGB}{52,152,219}
\definecolor{niceorange}{RGB}{243,156,18}
\definecolor{nicered}{RGB}{231,76,60}
\definecolor{niceyellow}{RGB}{241,196,15}
\definecolor{nicegrey}{RGB}{189,195,199}
\definecolor{nicepurple}{RGB}{155,89,182}

\begin{document}
\title{Providing Meaningful Data Summarizations Using Exemplar-based Clustering in Industry 4.0}
\titlerunning{Providing Meaningful Data Summarizations in Industry 4.0}

\author{Philipp-Jan Honysz\inst{1}\orcidID{0000-0003-1900-5285} \and
Alexander Schulze-Struchtrup\inst{2} \and
Sebastian Buschjäger\inst{1}\orcidID{0000-0002-2780-3618} \and
Katharina Morik\inst{1}\orcidID{0000-0003-1153-5986}}

\authorrunning{Honysz et al.}

\institute{Artificial Intelligence Unit, TU Dortmund University, Dortmund, Germany \and
Weppler Filter GmbH, Oberursel, Germany}

\maketitle              
\begin{abstract}
Data summarizations are a valuable tool to derive knowledge from large data streams and have proven their usefulness in a great number of applications. Summaries can be found by optimizing submodular functions. These functions map subsets of data to real values, which indicate their ''representativeness'' and which should be maximized to find a diverse summary of the underlying data. In this paper, we studied Exemplar-based clustering as a submodular function and provide a GPU algorithm to cope with its high computational complexity. We show, that our GPU implementation provides speedups of up to 72x using single-precision and up to 452x using half-precision computation compared to conventional CPU algorithms. We also show, that the GPU algorithm not only provides remarkable runtime benefits with workstation-grade GPUs but also with low-power embedded computation units for which speedups of up to 35x are possible. Furthermore, we apply our algorithm to real-world data from injection molding manufacturing processes and discuss how found summaries help with steering this specific process to cut costs and reduce the manufacturing of bad parts. Beyond pure speedup considerations, we show, that our approach can provide summaries within reasonable time frames for this kind of industrial, real-world data.
\end{abstract}

\section{Introduction}
\label{sec:introduction}

The developments of the last few years have made it possible to observe more and more data of various kinds in an ever shorter time. This is especially true for industrial processes, which are characterized by a large number of sensors and actuators to observe. Producers increasingly intend to use data from their machines to reduce the production of bad parts and to establish a consistent level of quality. However in these (industrial) environments, it is troublesome to identify those few data points, that matter to accomplish an effective and targeted process control. 

One way to tackle this problem is to derive data summarizations, which contain a selection of representative observations from a given dataset. Technically, this can be accomplished by optimizing submodular functions, which numerically assess the representativity of a particular subset of data. An optimization problem is then imposed by seeking a rather compact set, which maximizes the function and thus representativity. The Informative Vector Machine (IVM) assesses the representativity of a given set by considering Gram matrices consisting of Mercer kernel values, which appropriately need to be scaled for a given dataset. The IVM function evaluation is of low computational complexity but requires a careful and tailored selection of kernel and kernel scaling as this crucially determines the quality of the found summary. Exemplar-based clustering (EBC) is a different submodular function, which does not require fine-tuned hyperparameters but rather considers distances of candidate vectors w.r.t. the whole dataset to implicitly determine representativity. However, EBC suffers from high computational complexity.

In this work, we present a novel GPU algorithm, which accelerates exemplar-based clustering function evaluation. However, we not only provide a simple acceleration of distance computations but instead tailor the algorithm to be fast, when it is used together with submodular function optimizers. Our contributions are as follows:
\begin{itemize}
    \item We present the first algorithm to evaluate Exemplar-based clustering on GPUs and discuss, how this procedure exploits hardware features like shared memory and coalesced access to minimize runtime. 
	\item We conduct a series of experiments to compare different CPU implementations with our GPU algorithm and determine the possible benefits w.r.t. the achievable run-time and speedup. Our experiments, both focus on high-end devices and explicit low-power equipment. 
	\item We show, how found data summarizations can be used to steer production processes effectively. To accomplish this, we conduct a case study with data from injection molding machinery. Moreover we show, that these data summarizations can be provided within a reasonable time frame.
\end{itemize}

This paper is organized as the following: In section \ref{sec:related_work} we give a short overview on related work. In section \ref{sec:submodular_functions}, we formally establish submodular functions and the important cardinality-constrained optimization problem. Furthermore, we introduce the Greedy optimizer. In section \ref{sec:exemplar_based_clustering} we will establish the submodular function of Exemplar-based clustering, briefly explain how it measures representativity, discuss, how an implementation for CPUs might look like and which acceleration possibilities are feasible, before introducing our GPU algorithm. In section \ref{sec:experiments} we present our experiments and the achieved results. In section \ref{sec:case_study} we show how EBC-based data summarizations can be used to adjust injection molding machinery to reduce the production of bad parts and to cut costs. In section \ref{sec:conclusion} we summarize our work and give an outlook on how this work might be refined.

\section{Related Work}
\label{sec:related_work}

Submodular functions have proven their relevance to practical problems in several use cases: They have already been employed to detect events in activity networks \cite{Rozenshtein/etal/2014a} or to place sensors effectively \cite{Bellala/etal/2012a}. Submodular functions are also well-known tools to summarize data sets or data streams and select representative observations from them. Their use has been proposed to summarize video data \cite{Gygli/etal/2015a} or text corpora \cite{Lin/etal/2009a}. The optimization of submodular functions is a theoretically well-studied subject \cite{Nemhauser/Wolsey/78a,Nemhauser/etal/78a} with algorithms like Three Sieves \cite{Buschjaeger/etal/2020b} and Sieve Streaming \cite{Badanidiyuru/etal/2014a} providing strong guarantees towards the achievable function value and the ability to optimize with streaming data. To the best of our knowledge, the acceleration of Exemplar-based clustering as a submodular function has not been discussed so far. The acceleration of clustering, however, has been (among others) discussed for density-based clustering \cite{Andrade/etal/2013a}, $k$-means clustering \cite{Bhimani/etal/2015a}, $k$-medoids, and self-organizing maps \cite{Kohlhoff/etal/2011a} but not for Exemplar-based clustering. Moreover, to the best of our knowledge, data summaries using submodular functions have not been applied to industrial data for process control yet.

\section{Submodular Functions}
\label{sec:submodular_functions}

Submodular functions are set functions, which introduce a property of diminishing returns. To provide an adequate description of this characteristic, the \textit{discrete derivative} is being formalized as follows \cite{Krause/Golovin/2012a}:
\begin{definition}[Discrete derivative]
Let $V$ be a finite set (the ``ground set''), $f \colon \mathcal{P}(V) \rightarrow \mathbb{R}$, $S \subseteq V$ and $e \in V$. Then, $\Delta_f(e \mid S) = f(S \cup \left\lbrace e \right\rbrace) - f(S)$ represents the \textit{discrete derivative} of $f$ and the set $S$ w.r.t. the observation $e$.
\end{definition}
This definition directly leads to the formalization of submodular functions \cite{Krause/Golovin/2012a}:
\begin{definition}[Submodular function]
A function $f\colon \mathcal{P}(V) \rightarrow \mathbb{R}$ is \textit{submodular}, iff for all $A \subseteq B \subseteq V$ and $e \in V \setminus B$ the following condition holds:
\begin{align}
\Delta_f(e \mid A) \geq \Delta_f(e \mid B)
\end{align}
\end{definition}
This definition does not enforce monotonicity, which, however, is often required for optimization with strong guarantees. Hence, it is established in the following way \cite{Krause/Golovin/2012a}:
\begin{definition}[Monotonic submodular functions]
A submodular function $f$ is \textit{monotone}, iff for all $A \subseteq B \subseteq V$ it holds, that $f(A) \leq f(B)$.
\end{definition}

The problem of submodular function maximization is usually regarded in its cardinality constrained formulation, which we are also discussing here:
\begin{align}
S_k^* = \underset{S \subseteq V, |S| \leq k}{\max} f(S)
\label{eqn:submodular_function_optimization_cardinality_constraint}
\end{align}
Finding the global optimal solution to problem \ref{eqn:submodular_function_optimization_cardinality_constraint} is NP-hard \cite{Krause/Golovin/2012a} as this would require evaluating an exponential number of sets for their function value. 

It is well-known that the optimal function value with a polynominal number of function evaluations is $(1 - e^{-1}) \approx 63,21\%$ \cite{Nemhauser/Wolsey/78a}. Interestingly, this approximation is achieved with a simple Greedy algorithm \cite{Nemhauser/etal/78a}.

\section{Exemplar-based Clustering}
\label{sec:exemplar_based_clustering}

Exemplar-based clustering is strongly related to $k$-medoids clustering and its accompanying loss function, which is subject to minimization and may be defined as follows \cite{Gomes/Krause/2010a}:
\begin{definition}[$k$-medoids loss]
Let $\mathcal{X}$ be some data space, $V \subseteq \mathcal{X}$ a finite set, $d: \mathcal{X} \times \mathcal{X} \rightarrow \mathbb{R}^+$ some distance function and $S \subseteq V$. The $k$-medoids loss is then defined by
\begin{align}
L(S) = |V|^{-1} \sum_{\vec{v} \in V} \min_{\vec{s} \in S} d(\vec{v}, \vec{s})
\label{eqn:k_medoids_loss}
\end{align}
\end{definition}

One can formulate that problem as a monotone submodular function \cite{Gomes/Krause/2010a}.
\begin{definition}[Exemplar-based clustering]
Let $L: \mathcal{X} \rightarrow \mathbb{R}^+$ be the $k$-medoids loss function. Furthermore, let $\vec{e}_0 \in \mathcal{X}$ be some auxiliary vector, e.g. the all-zero vector $\vec{e}_o = (0,\dots,0)^T$. Then, \textit{exemplar-based clustering} may be defined as a monotone submodular function, as follows:
\begin{align}
f(S) = L(\left\lbrace \vec{e}_0 \right\rbrace) - L(S \cup \left\lbrace \vec{e}_0 \right\rbrace)
\label{eqn:exemplar_based_clustering}
\end{align} 
\end{definition}

Equation \ref{eqn:exemplar_based_clustering} measures the representativity of a selected set $S$ by computing the mean distance of all vectors to their nearest representative: If these distances are very high on average -- which is the case, when data vectors are very dissimilar to their nearest representative -- then, the $k$-medoids loss becomes larger, indicating poor representativeness.

\subsection{Evaluation on CPUs}
\label{sec:exem_eval_cpu}

To evaluate equation \ref{eqn:exemplar_based_clustering} one can consider a simple procedure as depicted in algorithm \ref{alg:exemplar_based_clustering_cpu}. This na\"ive implementation has a runtime complexity of $\mathcal{O}(|V| \cdot |S|)$, which is problematic for large $V$ and $S$. It is natural to consider index-structures to accelerate heavily distance-dependent algorithms. In fact, algorithm \ref{alg:exemplar_based_clustering_cpu} is a brute-force method, which may be characterized by a nearest neighbor-queries. Index structures, like $k$-$d$-trees, require that they are built upon some subset of the data space, which then can be queried for nearest neighbors \cite{Yianilos/1993a}. For equation \ref{eqn:k_medoids_loss} this would require establishing an index on the set $S$, which during optimization changes for \textit{every} function evaluation. Hence, we do not consider the use of index structures.

\begin{algorithm}[t]
   \caption{Exemplar-based clustering (CPU)}
   \label{alg:exemplar_based_clustering_cpu}
	\begin{algorithmic}
		\State \textbf{Input}: Datasets $V$ and $S \subseteq V$, Dissimilarity function $d$
		\Function{$L$}{$V, S$}
			\ForAll{$\vec{v}_i \in V$}
				\State $t \leftarrow $\texttt{FLT\_MAX}
				\ForAll{$\vec{s} \in S$}
					\State $t \leftarrow min(t, d(\vec{s}, \vec{v}_i))$
				\EndFor
				\State $\Sigma_i \leftarrow t$
			\EndFor
			\State $\sigma \leftarrow$ \textbf{reduce} $\Sigma$ \textbf{by} sum
			\State \textbf{return} $|V|^{-1}\sigma$
		\EndFunction
		\State \textbf{return} $L(V, \left\lbrace \vec{e}_0 \right\rbrace) - L(V, S \cup \left\lbrace \vec{e}_0 \right\rbrace)$
	\end{algorithmic}
\end{algorithm}

It is possible to accelerate algorithm \ref{alg:exemplar_based_clustering_cpu} by parallelizing the outer loop, which computes partial sums for every $\vec{v}_i \in V$ and therefore leads to $\mathcal{O}(|V|)$ computational tasks. However it has to be noted, that a large group of optimizers, like Greedy or Sieve Streaming \cite{Badanidiyuru/etal/2014a} do not consider a single set $S$ for evaluation but multiple sets $S_\text{multi} = \left\lbrace S_1, \dots, S_l \right\rbrace$ in every optimization step. Consider the Greedy algorithm: In every optimization step the procedure decides, which not yet selected item from the ground set leads to the greatest marginal gain in terms of function value. Given the currently optimal set $S_i$ and a set of candidate items $C = \left\lbrace \vec{c}_1, \dots, \vec{c}_m \right\rbrace$, which might be selected in the current optimization step, this leads to $S_\text{multi} = \left\lbrace S_i \cup \left\lbrace \vec{c}_1 \right\rbrace, \dots, S_i \cup \left\lbrace \vec{c}_m \right\rbrace \right\rbrace$, which needs to be evaluated for their respective function values. While $\mathcal{O}(|V|)$ computational tasks already do not seem to be feasible for contemporary multi-core systems, the multiset parallelized problem imposes $\mathcal{O}(|V| \cdot |S_\text{multi}|)$ tasks, which even more stresses the need for hardware that is better suited to solve that massive number of independent computational problems. This is especially true, since $|C| \approx |V|$ during Greedy optimization. 

\subsection{Evaluation on GPUs}

GPUs are a very popular choice to accelerate massively parallel tasks due to its many-core architecture and its general availability to the wide public. Hence, we choose the GPU to accelerate EBC and discuss an appropriate implementation. Although different models for general purpose GPU (GPGPU)-programming exist, we will focus on the CUDA framework throughout this work.

While it is possible to organize threads on CPUs in virtually arbitrary ways, for GPUs we have to map the specific job into a grid-block structure. Every \textit{grid} may have up to three dimensions, which in turn contains \textit{blocks}. Likewise, every block may also have up to three dimensions and consists of no more than 1024 threads. It is important to note, that threads are issued in \textit{warps}, which represent a group of usually 32 threads that are launched together at the same address and process the same instruction. Every thread from a single block has access to comparably small \textit{shared memory}, which in contrast to \textit{global memory} represents on-chip memory. Hence, accesses to shared memory are relatively fast.

We establish a formal framework to discuss the mapping into the grid-block structure, as follows: Let $C = (D_g, D_b)$ be a specific \textit{kernel configuration} to solve a particular problem, with $D_g = (g_x, g_y, g_z)$ and $D_b = (b_x, b_y, b_z)$ the dimensioning of the grid and the block respectively. Every thread knows, which position $P_t = (t_x^*, t_y^*, t_z^*)$ it acquires in a specific block $P_b = (b_x^*, b_y^*, b_z^*)$. This information is essential, since every individual thread derives from it which task it has to conduct.

\subsubsection{Algorithm}

For our GPU implementation we consider equation \ref{eqn:exemplar_based_clustering} again: It can be seen, that the expression $L(\left\lbrace \vec{e}_0 \right\rbrace)$ is independent of the given set $S$ to evaluate. Since that sub-expression is $\mathcal{O}(|V|)$ in terms of time complexity, we do not consider any acceleration and compute that term conventionally, which makes the resulting value available to \textit{all} subsequent computations. 

We will focus on the expression $L(S \cup \left\lbrace \vec{e}_0 \right\rbrace)$ from now on: To ensure, that we can place as many independent tasks on the GPU as possible, we consider a decomposition of the aforementioned sub-expression, as follows:
\begin{align}
		\label{eqn:loss_decomposed}
        L_{\vec{v}_i} (S) = |V|^{-1} \min_{\vec{s} \in S} d(\vec{v}_i, \vec{s}) \\
		\label{eqn:loss_decomposed_summarized}
        L(S) = L_{\vec{v}_1} (S) + \cdots + L_{\vec{v}_n} (S)
\end{align}
Using that decomposition and the set of sets to evaluate $S_\text{multi}$ we can construct a work matrix $\mathbf{W}$, which constitutes the foundation to design the grid-block structure from:
\begin{align}
        \mathbf{W} = \begin{pmatrix}
        L_{\vec{v}_1}(S_1) & L_{\vec{v}_2}(S_1) & \cdots & L_{\vec{v}_n}(S_1) \\ 
        \vdots & \vdots  & \ddots &  \vdots \\ 
        L_{\vec{v}_1}(S_l) & L_{\vec{v}_2}(S_l) & \cdots &  L_{\vec{v}_n}(S_l)
        \end{pmatrix}
		\label{eq:work_matrix}
\end{align}
It can be seen, that a reduction of equation \ref{eq:work_matrix} by sum in column direction leads back to equation \ref{eqn:loss_decomposed_summarized}, which is the result we are ultimately looking for. We design our procedure so that every GPU thread is assigned to a single cell from $\mathbf{W}$. 

As algorithm \ref{alg:exemplar_based_clustering_cpu} already suggests, many memory accesses are made w.r.t. the elements from $V$. Therefore, this should be performed as fast as possible. Hence, we load vectors $\vec{v}_i \in V$ from global memory into shared memory, that serves as a low latency, user-managed cache. Because shared memory is block exclusive, we strive to put as many threads $L_{\vec{v}_i}$ into one block as possible. From the perspective of equation \ref{eq:work_matrix}, that means that we want to maximize the block dimensions in row direction. To accomplish this, we have to provide an appropriate block dimensioning $D_b$, which considers both the restrictions of the programming model (max. 1024 threads per block) and the number of bytes $\beta$ we may allocate per block from shared memory. Let $\gamma$ be the number of bytes every $\vec{v}_i \in V$ requires to be stored, then we can determine $D_b = (b_x, b_y, b_z = 1)$ to be as follows:
\begin{align*}
b_x = \min \left\lbrace \left\lfloor \frac{1024}{b_Y} \right\rfloor, \left\lfloor \frac{\beta}{\gamma} \right\rfloor \right\rbrace \;
b_y = \min \left\lbrace 1024, |S_{\text{multi}}| \right\rbrace
\end{align*}
It can be seen that $b_y$ represents the maximization we have discussed before. Conversely, $b_x$ is constructed by instantiating as many threads as we have still left after considering $b_y$. Additionally, if the block is growing in column direction, we are taking new ground vectors into account. Hence, we also have to take care, that we do not exceed the per-block shared memory size limitation $\beta$. 

To provide a complete kernel configuration $C = (D_g, D_b)$, we now have to specify a grid dimensioning $D_g$. From a given block dimensioning $D_b$ we construct $D_g = (g_x, g_y, g_z = 1)$ such that every cell from $\mathbf{W}$ is being computed:
\begin{align}
g_x = \left\lceil \frac{|V|}{b_X} \right\rceil \; g_y = \left\lceil \frac{|S_{\text{multi}}|}{b_Y} \right\rceil
\end{align}

\begin{algorithm}[t]
        \caption{Exemplar-based clustering (GPU)}
        \begin{algorithmic}
				\Require{dissimilarity function $d$}
                \State{$i \leftarrow b_x \cdot b_x^* + t_x^*, j \leftarrow b_y \cdot b_y^* + t_y^*$}
                \If{$t_y^* = 0$}
                        \State{\textbf{load} $\vec{v}_i \in V$ \textbf{from} global memory \textbf{into} shared memory}
                \EndIf
                \State{\textbf{synchronize} threads \textbf{in} block}
                \State{$d_\text{min} \leftarrow $ \texttt{FP\_MAX}}
                \ForAll{$\vec{s}_i \in S_j$}
                        \State{$d_\text{min} \leftarrow \min(d_\text{min}, d(\vec{s}_i, \vec{v}_i))$}
                \EndFor
                \State{$W_{j, i} = \frac{d_\text{min}}{|V|}$}
        \end{algorithmic}
        \label{alg:exemplar_based_clustering_gpu}
\end{algorithm}

As we have established the kernel configuration, we will now focus on the concrete kernel, i.e. the computational routine to calculate every cell from $\mathbf{W}$. The kernel is depicted in algorithm \ref{alg:exemplar_based_clustering_gpu}: First, every thread determines to which cell of $\mathbf{W}$ it belongs to by calculating the $i$ and $j$ index. Then, every first thread in $y$ direction loads the accompanying vector from global into shared memory. The following lines compute equation \ref{eqn:loss_decomposed}, whereby the final result is ultimately written back to the work matrix, which resides in global memory. The resulting work matrix is then reduced in a row-wise fashion on the GPU by calculating $\mathbf{W} \cdot \vec{1}$, which delivers the final result for every $S \in S_\text{multi}$.

\subsubsection{Memory Layout}

So far we are assuming that our computational payload (i.e. $V$ and $S_\text{multi}$) \textit{somehow} has been copied and stored in global memory. To ensure efficiency, we have to copy the payload in as few transactions as possible. Furthermore, we have to present the data in an advantageous way to the GPU. Let us now discuss this in more detail.

CPU and GPU memory do not differ much when it comes to accessing techniques. Hence, vectors can easily be stored by successively writing its contents to memory. However, matrices represent \textit{complex} data structures since they require two-dimensional addressing across rows and columns. Therefore, we need to map matrices to ``ordinary'' vectors. Let us call this process \textit{vectorization}. For our algorithm, we store the $\mathbf{V}$ matrix in a column-wise fashion and copy it in a single memory transaction to the GPU. Since the ground matrix never changes between different function evaluations it is copied to the GPU's global memory on algorithm initialization. We do not consider further optimization, because no further memory accesses on the GPU, beyond loading vectors from global to shared memory, are conducted, as discussed above.

This is different when it comes to the set of evaluation matrices $S_\text{multi}$ since they are too large to be cached in shared memory. Therefore, access by global memory is mandatory and has to be conducted as efficiently as possible. In CUDA, global memory is divided into segments of fixed size, which usually is 32 bytes. Therefore a segment allows for up to four 64-bit floating-point numbers (FP64) or up to eight 32-bit floating-point numbers (FP32). If threads of a single warp access a particular memory segment, then these memory accesses become \textit{coalesced} into a single memory transaction, which is beneficial to runtime and data throughput. Conversely, if threads of a single warp access different memory segments then more memory transactions are needed to access the same data.

\begin{figure}[t]
	\centering
	\resizebox{0.91\columnwidth}{!}{%
		\begin{tikzpicture}[
			matrixstyle/.style={matrix of nodes, nodes in empty cells, column sep=-\pgflinewidth, row sep=-\pgflinewidth, nodes={inner sep=0mm,outer sep=0pt, minimum size=5mm, text height=\ht\strutbox,text depth=\dp\strutbox, draw}}, 
			brace/.style={decoration={brace}, decorate},
   			position label/.style={above = 2pt, text height = 2ex, text depth = 1ex}
			]
		
		\matrix (S1) at (0, 30mm) [matrixstyle, ampersand replacement=\&]{
			|[fill=nicegreen!50]| $\cdot$ \& |[fill=nicegreen!50]| $\cdot$ \& |[fill=nicegreen!50]| $\cdot$ \& |[fill=nicegreen!50]| $\cdot$ \\
			|[fill=nicegreen!50]| $\cdot$ \& |[fill=nicegreen!50]| $\cdot$ \& |[fill=nicegreen!50]| $\cdot$ \& |[fill=nicegreen!50]| $\cdot$ \\
		};
	
		\matrix (S2) at (25mm, 30mm) [matrixstyle, ampersand replacement=\&]{
			|[fill=niceblue!50]| $\cdot$ \& |[fill=niceblue!50]| $\cdot$ \& |[fill=niceblue!50]|$\cdot$ \\
			|[fill=niceblue!50]| $\cdot$ \& |[fill=niceblue!50]| $\cdot$ \& |[fill=niceblue!50]|$\cdot$ \\
		};
	
		\matrix (S3) at (52.5mm, 30mm) [matrixstyle, ampersand replacement=\&]{
			|[fill=niceyellow!50]| $\cdot$ \& |[fill=niceyellow!50]| $\cdot$ \& |[fill=niceyellow!50]| $\cdot$ \& |[fill=niceyellow!50]| $\cdot$ \& |[fill=niceyellow!50]| $\cdot$ \\
			|[fill=niceyellow!50]| $\cdot$ \& |[fill=niceyellow!50]| $\cdot$ \& |[fill=niceyellow!50]| $\cdot$ \& |[fill=niceyellow!50]| $\cdot$ \& |[fill=niceyellow!50]| $\cdot$ \\
		};

		\node[] (S3DBrTop) at ($(S3-1-5.north east)+(0.3,-0.2)$) {};
		\node[] (S3DBrBottom) at ($(S3-2-5.south east)+(0.3,0.1)$) {};
		
		\node[] (S1BrLeft) at ($(S1-1-1.north west)+(0.0,0.325)$) {};
		\node[] (S1BrRight) at ($(S1-1-4.north east)+(0.0,0.325)$) {};
		
		\node[] (S2BrLeft) at ($(S2-1-1.north west)+(0.0,0.325)$) {};
		\node[] (S2BrRight) at ($(S2-1-3.north east)+(0.0,0.325)$) {};

		\node[] (S3BrLeft) at ($(S3-1-1.north west)+(0.0,0.325)$) {};
		\node[] (S3BrRight) at ($(S3-1-5.north east)+(0.0,0.325)$) {};

		\draw [brace] (S3DBrTop.north) -- node [pos=0.5, xshift=3.5mm] {$d$} (S3DBrBottom.south);
		\draw [brace] (S1BrLeft.south) -- node [position label, pos=0.5] {$|S_1|$} (S1BrRight.south);
		\draw [brace] (S2BrLeft.south) -- node [position label, pos=0.5] {$|S_2|$} (S2BrRight.south);
		\draw [brace] (S3BrLeft.south) -- node [position label, pos=0.5] {$|S_3|$} (S3BrRight.south);
		
		\matrix (SMat) at (27.5mm, 12.5mm) [matrixstyle, ampersand replacement=\&]{
			|[fill=nicegreen!50]| $\cdot$ \&|[fill=niceblue!50]| $\cdot$ \& |[fill=niceyellow!50]| $\cdot$ \& 
			|[fill=nicegreen!50]| $\cdot$ \&|[fill=niceblue!50]| $\cdot$ \& |[fill=niceyellow!50]| $\cdot$ \& 
			|[fill=nicegreen!50]| $\cdot$ \&|[fill=niceblue!50]| $\cdot$ \& |[fill=niceyellow!50]| $\cdot$ \&
			|[fill=nicegreen!50]| $\cdot$ \&                     $\cdot$ \& |[fill=niceyellow!50]| $\cdot$ \&
                                  $\cdot$ \&                     $\cdot$ \& |[fill=niceyellow!50]| $\cdot$ \\
			|[fill=nicegreen!50]| $\cdot$ \&|[fill=niceblue!50]| $\cdot$ \& |[fill=niceyellow!50]| $\cdot$ \& 
			|[fill=nicegreen!50]| $\cdot$ \&|[fill=niceblue!50]| $\cdot$ \& |[fill=niceyellow!50]| $\cdot$ \& 
			|[fill=nicegreen!50]| $\cdot$ \&|[fill=niceblue!50]| $\cdot$ \& |[fill=niceyellow!50]| $\cdot$ \&
			|[fill=nicegreen!50]| $\cdot$ \&                     $\cdot$ \& |[fill=niceyellow!50]| $\cdot$ \&
                      $\cdot$ \&                     $\cdot$ \& |[fill=niceyellow!50]| $\cdot$ \\
		};

		\node[xshift=2mm, yshift=-5mm, position label] (SMatLabel) at (SMat.east) {$\mathbf{S}$};
		
		\path[draw=black!10, ->] (S1-2-2.south) to [out = 270, in = 90, looseness = 0.5] (SMat-1-4.north);
		\path[draw=black!10, ->] (S1-2-3.south) to [out = 270, in = 90, looseness = 0.5] (SMat-1-7.north);
		\path[draw=black!10, ->] (S1-2-4.south) to [out = 270, in = 90, looseness = 0.5] (SMat-1-10.north);
		
		\path[draw=black!10, ->] (S2-2-2.south) to [out = 270, in = 90, looseness = 0.5] (SMat-1-5.north);
		\path[draw=black!10, ->] (S2-2-3.south) to [out = 270, in = 90, looseness = 0.5] (SMat-1-8.north);
	
		\path[draw=black!10, ->] (S3-2-2.south) to [out = 270, in = 90, looseness = 0.5] (SMat-1-6.north);
		\path[draw=black!10, ->] (S3-2-3.south) to [out = 270, in = 90, looseness = 0.5] (SMat-1-9.north);
		\path[draw=black!10, ->] (S3-2-4.south) to [out = 270, in = 90, looseness = 0.5] (SMat-1-12.north);
		\path[draw=black!10, ->] (S3-2-5.south) to [out = 270, in = 90, looseness = 0.5] (SMat-1-15.north);
		\draw[->] (S1-2-1.south) -- (SMat-1-1.north);
		\draw[->] (S2-2-1.south) to [out = 270, in = 90, looseness = 0.5] (SMat-1-2.north);
		\draw[->] (S3-2-1.south) to [out = 270, in = 90, looseness = 0.3] (SMat-1-3.north);
		
		\matrix (SMatVec) at (27.5mm, -2.5mm) [matrixstyle, ampersand replacement=\&]{
			|[fill=nicegreen!50]| $\cdot$ \&|[fill=niceblue!50]| $\cdot$ \& |[fill=niceyellow!50]| $\cdot$ \& 
			|[fill=nicegreen!50]| $\cdot$ \&|[fill=niceblue!50]| $\cdot$ \& |[fill=niceyellow!50]| $\cdot$ \& 
			|[fill=nicegreen!50]| $\cdot$ \&|[fill=niceblue!50]| $\cdot$ \& |[fill=niceyellow!50]| $\cdot$ \&
			|[fill=nicegreen!50]| $\cdot$ \&                     $\cdot$ \& |[fill=niceyellow!50]| $\cdot$ \&
                                  $\cdot$ \&                     $\cdot$ \& |[fill=niceyellow!50]| $\cdot$ \&
			|[fill=nicegreen!50]| $\cdot$ \&|[fill=niceblue!50]| $\cdot$ \& |[fill=niceyellow!50]| $\cdot$ \& 
			|[fill=nicegreen!50]| $\cdot$ \&|[fill=niceblue!50]| $\cdot$ \& |[fill=niceyellow!50]| $\cdot$ \& 
			|[fill=nicegreen!50]| $\cdot$ \&|[fill=niceblue!50]| $\cdot$ \& |[fill=niceyellow!50]| $\cdot$ \&
			|[fill=nicegreen!50]| $\cdot$ \&                     $\cdot$ \& |[fill=niceyellow!50]| $\cdot$ \&
                      $\cdot$ \&                     $\cdot$ \& |[fill=niceyellow!50]| $\cdot$ \\
		};

		\path[draw=black, ->] (SMat-2-7.south) -- node [text width=2.5cm,midway,right] {Vectorization} ($(SMat-2-7.south)-(0.0,0.75)$);
	
		\node[] (T1a) at ($(SMatVec-1-1.south west)-(-0.25,0.75)$) {$t_1$};
		\node[] (T2a) at ($(SMatVec-1-2.south west)-(-0.25,0.75)$) {$t_2$};
		\node[] (T3a) at ($(SMatVec-1-3.south west)-(-0.25,0.75)$) {$t_3$};
		\node[] (T1b) at ($(SMatVec-1-4.south west)-(-0.25,0.75)$) {$t_1$};
		\node[] (T2b) at ($(SMatVec-1-5.south west)-(-0.25,0.75)$) {$t_2$};
		\node[] (T3b) at ($(SMatVec-1-6.south west)-(-0.25,0.75)$) {$t_3$};
		\node[] (Tdots) at ($(SMatVec-1-7.south west)-(-0.375,0.75)$) {$\cdots$};
		
		\draw[->] (T1a.north)  -- (SMatVec-1-1.south);
		\draw[->] (T2a.north)  -- (SMatVec-1-2.south);
		\draw[->] (T3a.north)  -- (SMatVec-1-3.south);
		\draw[->] (T1b.north)  -- (SMatVec-1-4.south);
		\draw[->] (T2b.north)  -- (SMatVec-1-5.south);
		\draw[->] (T3b.north)  -- (SMatVec-1-6.south);
	\end{tikzpicture}
	}
	\caption{Three evaluation matrices $\mathbf{S}_1$, $\mathbf{S}_2$ and $\mathbf{S}_3$ with four, three and five elements respectively and a dimensionality of $d=2$ are being processed into a single matrix $\mathbf{S}$. Subsequently, the resulting matrix is being \textit{vectorized} in a row-wise fashion. The threads $t_1$, $t_2$ and $t_3$ are assigned to the evaluation matrices $\mathbf{S}_1$, $\mathbf{S}_2$ and $\mathbf{S}_3$ and successively access the information of vectorized matrix, which leads to coalesced access and to as few memory transactions as possible.}
	\label{fig:ExemGPU_BuildSummaryMatrix}
\end{figure}
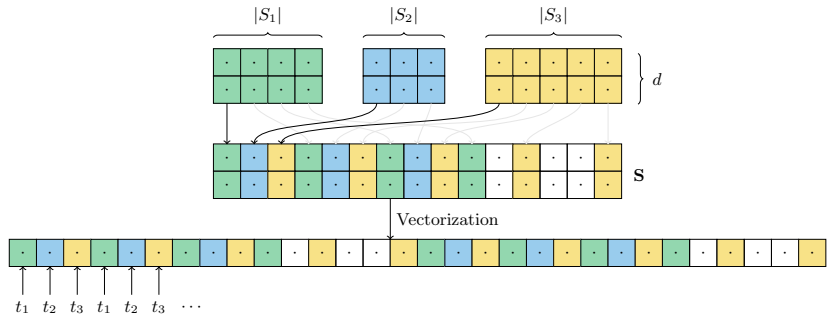

In addition to exploiting mentioned locality effects, we also want to copy the evaluation matrices in as few requests as possible. Copying computational payload is a relatively expensive operation and the bandwidth of the PCI-E connection (which usually serves as the link between CPU and GPU) is only exhausted when sufficiently large data is transmitted. Hence, we propose the following vectorization routine, which not only is aware of locality effects but also seeks to exhaust the bandwidth of the interconnect between CPU and GPU. 

To discuss our vectorization routine, we again take a look at algorithm \ref{alg:exemplar_based_clustering_gpu}. Especially the calculation of the dissimilarity value between two vectors $\vec{s}_i$ and $\vec{v}_i$ is of interest since this is the portion of the code, where data accesses happen. Most dissimilarity functions require to loop over the elements of the vectors to compare. Since $\vec{v}_i$ is already loaded into shared memory, we focus on the various memory accesses to $\vec{s}_i$ from different GPU kernel threads. As we are seeking to optimize accesses of threads of a particular warp, we can assume that every thread processes the same instruction. Let loading $\vec{s}_i[k] \in S_j$ from memory be now the instruction of interest, whereas $k$ indexes a particular dimension. From work matrix $\mathbf{W}$ we can assume, that $i$ and $k$ are equal within threads of the same warp (cf. equation \ref{eq:work_matrix}). To exploit coalesced memory access, we have to optimize loading vectors from $S_j$, whereas $j$ varies for different threads of the same warp and the same block. Hence, the data of the different sets $S_j$ for which $k$ and $i$ are equal must be stored sequentially in memory. From a technical point of view, this can be achieved by choosing an evaluation set $S_j$ in round robin-fashion and selecting the next, not yet processed vector from that set. The selected vector is then written to an evaluation set matrix $\mathbf{S}$, which contains the data of all evaluation sets. If all vectors from a chosen set have been written to the matrix, the entry in $\mathbf{S}$ simply remains empty.

\section{Experiments}
\label{sec:experiments}

\begin{figure}[t]
	\centering
	\resizebox{\textwidth}{!}{%
		\begin{tikzpicture}
			\begin{groupplot}[group style={group name=runtimeplot, group size=3 by 1, horizontal sep=1.5cm, vertical sep=1.5cm}]
				\def\myPlots{}
				\pgfplotsforeachungrouped \y in {FP32}{
					\pgfplotsforeachungrouped \x in {N, l, k}{
						\eappto\myPlots{
							\noexpand\nextgroupplot[xlabel=$\x$, ylabel={Runtime [s]}, title=Variation of $\x$, 
							legend style={at={($(0, 0)+(1cm,1cm)$)},legend columns=3,fill=none,draw=black,anchor=center,align=center}, legend to name=runtimeplotlegend,
							scaled x ticks=false, xticklabel style={
								/pgf/number format/fixed,
								/pgf/number format/precision=5
							},
							scaled y ticks=false, yticklabel style={
								/pgf/number format/fixed,
								/pgf/number format/precision=5
							}, y post scale=0.6, ymode=log]

							\noexpand\addplot[color=niceblue, mark=*, mark options={draw=black}] table[x=\x, y=RUNTIME_QUADRO, col sep=comma] {results/\x-\y-speedup.csv};
							\noexpand\addlegendentry{GPU (Quadro RTX 5000)}

							\noexpand\addplot[color=nicered, mark=*, mark options={draw=black}] table[x=\x, y=RUNTIME_INTEL_ST, col sep=comma] {results/\x-\y-speedup.csv};
							\noexpand\addlegendentry{Xeon W-2155 (Singlethread)}

							\noexpand\addplot[color=niceorange, mark=*, mark options={draw=black}] table[x=\x, y=RUNTIME_INTEL_MT, col sep=comma] {results/\x-\y-speedup.csv};
							\noexpand\addlegendentry{Xeon W-2155 (Multithread)}

							\noexpand\addplot[color=nicepurple, mark=*, mark options={draw=black}] table[x=\x, y=RUNTIME_A72_ST, col sep=comma] {results/\x-\y-speedup.csv};
							\noexpand\addlegendentry{Cortex-A72 (Singlethread)}

							\noexpand\addplot[color=nicegrey, mark=*, mark options={draw=black}] table[x=\x, y=RUNTIME_A72_MT, col sep=comma] {results/\x-\y-speedup.csv};
							\noexpand\addlegendentry{Cortex-A72 (Multithread)}

							\noexpand\addplot[color=nicegreen, mark=*, mark options={draw=black}] table[x=\x, y=RUNTIME_TX2, col sep=comma] {results/\x-\y-speedup.csv};
							\noexpand\addlegendentry{GPU (TX2)}

						}
					}
				}
				\myPlots
			\end{groupplot}

		\node[below] at (current bounding box.south) {\pgfplotslegendfromname{runtimeplotlegend};};
		\end{tikzpicture}
	}
	\caption{Runtimes of our GPU algorithm and of corresponding CPU algorithms.}
	\label{fig:runtimes_perfloss_variation}
\end{figure}
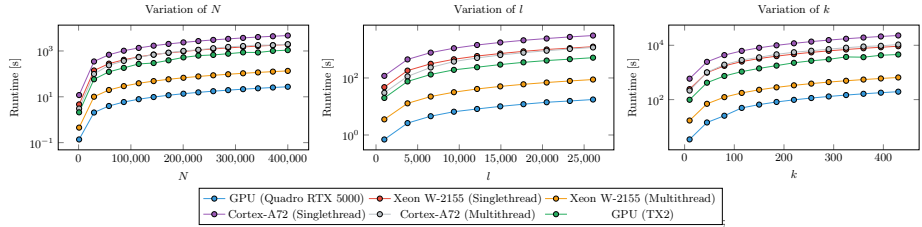

In this section, we experimentally evaluate our GPU algorithm and compare it to a CPU implementation, whereby two different variants will be considered: A single-threaded version of algorithm \ref{alg:exemplar_based_clustering_cpu}, which does not consider any parallelization at all, and a multi-threaded version, which runs the mentioned algorithm on different sets in parallel. Both CPU procedures make use of a SIMD strategy, to accomplish the sum reduction described in algorithm \ref{alg:exemplar_based_clustering_cpu}, whereby OpenMP is utilized to achieve that. The evaluation focuses on three research questions:
\begin{enumerate}
\item What is the impact of the number of vectors $N$ in the ground set $V$, the number of subsets $l$, and the number of vectors $k$ in every set $S \in S_\text{multi}$ on the real execution time?
\item How much does the runtime improve for embedded systems?
\item What is the impact of using half-precision (FP16) or single-precision (FP32) on wall-clock runtime?
\end{enumerate}
To discuss these questions adequately, we measure the wall-clock time it took to solve a particular problem, which consists of evaluating some ground set $V$ and a set of subsets $S_\text{multi}$. Every problem is randomly generated, whereby the data generation is \textit{not} part of the measured run-time. We chose to run the CPU experiments both on a workstation-grade Intel Xeon W-2155 and a low-power ARM Cortex-A72 built into a Raspberry Pi 4. For conducting the GPU experiments we chose to compare the performance of the algorithms by both employing a workstation-grade NVIDIA Quadro RTX 5000 as well as an embedded NVIDIA TX2 GPU. Furthermore, the squared Euclidean distance $d = \|\vec{x} - \vec{y}\|^2_2$ will be used as a dissimilarity measure between different observations for all our experiments.

\subsection{Variation of run-time-critical properties}
\label{sec:experiments_variation_runtime}

The runtime of the different algorithms is influenced by three factors: The number of observations in the ground set ($N$), the number of subsets to evaluate ($l$) and the number of observations in each of the $l$ subsets ($k$). For this discussion, we initially choose these parameters to be $N = 50000, l = 5000, k = 10$. Now, we consider three series of experiments, in which each of these parameters is varied while every other parameter is left at their initial value. The variations considered are $N \in \left\lbrace 1000, 29500, \dots, 400000\right\rbrace$, $l \in \left\lbrace 1000, 3785, \dots, 26070 \right\rbrace$ and $k \in \left\lbrace 10, 45, \dots, 430 \right\rbrace$. The dimensionality of every created observation is fixed to $100$. Furthermore, we only consider FP32 computation for this series of experiments since contemporary GPUs suffer in terms of runtime from requiring double precision floating point arithmetics. The results are depicted in figure \ref{fig:runtimes_perfloss_variation} and table \ref{tab:speedup}.

\begin{table}[t]
\caption{Achieved minimal, mean and maximal speedup of our GPU algorithm in 15 runs compared to a single-threaded (ST) and multi-threaded (MT) CPU implementation and different variations of $N$, $l$, or $k$. FP16-GPU speedups were computed from comparison with FP32-CPU wall-clock run-times.}
\centering
\input{results/speedup_agg.tex}
\label{tab:speedup}
\end{table}

We can derive, that our GPU algorithm is the fastest method to compute Exemplar-based clustering when FP32 is used. This remains true regardless of whether workstation or embedded hardware is used. Comparing the Quadro's runtime to the Xeon's runtime, the achievable speedup compared to the ST implementation ranges between 34x in the worst-case scenario and 72x in the best-case scenario. Conversely, the MT CPU implementation shows speedups ranging from 3.3x to 5.1x. Allowing FP16 computations on the Quadro GPU and comparing the wall-clock runtime to the Xeon's FP32 performance yields dramatic runtime improvements: Considering ST computations speedups ranging from 8.5x to 438.2x could be observed. Allowing MT computations on the CPU led to speedups ranging from 0.8x to 30.8x.

With respect to the embedded hardware, namely the NVIDIA TX2 GPU and the ARM Cortex-A72 CPU, our GPU algorithm yielded much less dramatic speedups but still noticeable runtime improvements: FP32 computations led to speedups ranging from 4.3x to 6x and from 1.5x to 2.7x for ST and MT computations respectively. As expected, FP16 operation on the embedded GPU yielded much better runtimes when compared to the ARM CPU: Compared to the ST CPU algorithm speedups between 5.1x and 35.5x were possible. Allowing MT computations on the CPU led to lower observed speedups ranging from 1.3x to 15.8x.

\section{Case Study}
\label{sec:case_study}

Injection molding is one of the most relevant ways to process polymer materials, with a market size expected to reach 470 billion USD by 2025 \cite{Watson2019}. Applications are found in nearly every part of daily life, including automotive interiors, exteriors and technical parts, medical technology items, office supplies as well as household appliances. The discontinuous process allows the efficient and highly reproducible production of molded parts with complex geometry meeting tight tolerances. This is achieved by several mainly subsequent process steps \cite{Osswald2008}: 

First, the usually granular material is plasticized, i.e., it is melted in a cylinder by frictional heat generated by a rotating screw as well as through additional heat transfer from heating elements. Once plasticized, the material is injected into a cavity, which has nearly the same shape as the final part, except it is slightly larger, since the thermoplastic material will shrink when cooling down to ambient temperature. To minimize shrinking, a holding pressure is applied while the part is in the cavity. Beginning with injection, the part in the cavity is cooling down and once a certain temperature is reached at which the part has the necessary rigidity, the mold opens, and the part is ejected before the cyclic process starts again.

\begin{figure}[t]
	\centering
	\resizebox{0.9\textwidth}{!}{%
		\begin{tikzpicture}
			\begin{groupplot}[group style={group name=runtimeplot, group size=3 by 1, horizontal sep=1.75cm, vertical sep=1.5cm}]
				\def\sgtPlots{}
				\pgfplotsforeachungrouped \y in {FP16, FP32}{
					\eappto\sgtPlots{
						\noexpand\nextgroupplot[xlabel=Summary Size ($k$), ylabel={Optimization Time $\left\lbrack s \right\rbrack$}, title=\y, 
						legend style={at={($(0, 0)+(1cm,1cm)$)},legend columns=4,fill=none,draw=black,anchor=center,align=center}, legend to name=sgtplotlegend,
						scaled x ticks=false, xticklabel style={
							/pgf/number format/fixed,
							/pgf/number format/precision=5
						},
						scaled y ticks=false, yticklabel style={
							/pgf/number format/fixed,
							/pgf/number format/precision=5
						}, y post scale=0.5, ymode=log]

						\noexpand\addplot[color=niceblue, mark=*, mark options={draw=black}] table[x=K, y=TIME_SECONDS, col sep=comma, discard if not={OPTIMIZER}{Greedy}] {results/summary_generation_times_\y_rtx5000.csv};
						\noexpand\addlegendentry{Quadro RTX 5000 (Greedy)}

						\noexpand\addplot[color=nicered, mark=*, mark options={draw=black}] table[x=K, y=TIME_SECONDS, col sep=comma, discard if not={OPTIMIZER}{Three Sieves}] {results/summary_generation_times_\y_rtx5000.csv};
						\noexpand\addlegendentry{Quadro RTX 5000 (Three Sieves)}

						\noexpand\addplot[color=niceorange, mark=*, mark options={draw=black}] table[x=K, y=TIME_SECONDS, col sep=comma, discard if not={OPTIMIZER}{Greedy}] {results/summary_generation_times_\y_tx2.csv};
						\noexpand\addlegendentry{Jetson TX2 (Greedy)}

						\noexpand\addplot[color=nicepurple, mark=*, mark options={draw=black}] table[x=K, y=TIME_SECONDS, col sep=comma, discard if not={OPTIMIZER}{Three Sieves}] {results/summary_generation_times_\y_tx2.csv};
						\noexpand\addlegendentry{Jetson TX2 (Three Sieves)}

					}
				}
				\sgtPlots
			\end{groupplot}

		\node[below] at (current bounding box.south) {\pgfplotslegendfromname{sgtplotlegend};};

		\end{tikzpicture}
	}
	\caption{Required optimization time to provide a summary of size $k$ from $N=1000$ time series with a dimensionality of $d=3524$. Submodular function optimization is conducted by employing Greedy and Three Sieves \cite{Buschjaeger/etal/2020b}. It can be seen, that rather large summarization can mostly be provided in reasonable time frames.}
	\label{fig:summary_generation_times}
\end{figure}
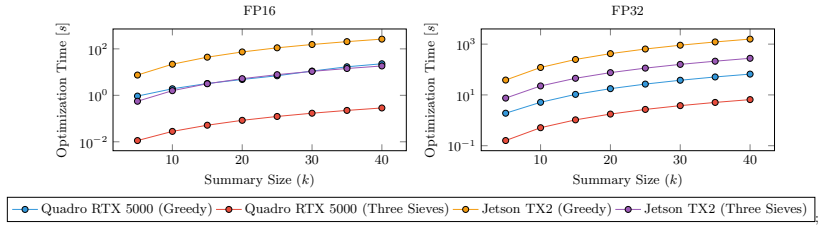

\subsubsection{Motivation}

In injection molding, the optimal process state is a stable process which is in thermal equilibrium without systematic external influences on process and molded part quality. Unfortunately, during real world injection molding production, this is rarely the case: When starting an IMM, the process is not yet in equilibrium, then for a period there may be stable process which is however interrupted by planned and unplanned downtimes of varying lengths. Also, material properties may change over time due to batch fluctuations etc., and changes of the machine setting parameters may become necessary to compensate external influences on the process. To evaluate how these process variations manifest in the process data and how they can be concentrated by data summarizations, five different process states have been induced for two molded parts, namely a plate and a cover: start-up, stable process, downtimes, regrind material as well as design of experiments (DOE).

Each of the ten datasets contains 1000 examples, except for the DOEs containing 860. In the start-up process, the IMM is started once the machine has reached a minimal necessary temperature, still being far from thermal equilibrium. During the stable process, no changes to the operation point are made and no external influence exist. In the downtimes process, we stopped the machine every 100 cycles for different amounts of time typical for real production. In the regrind material process, different amounts of regrind material are added to the granule, going from 0 \% to 100 \% in five steps. The DOE is a central composite design with star points and central point, yielding a total of 43 different machine settings.
Over the last decades, accuracy and reproducibility of injection molding machines (IMM) have increased significantly, so the potential for improvements in this area is considered exhausted. Still there are influences on the injection molding process, such as variations in the material properties or environmental conditions, which may negatively affect the quality of the molded parts. To address these issues, the research focus over the last few years has moved to improving control techniques \cite{Pillwein2011,Wang2013} and analyzing process data \cite{Schiffers/2018a,Struchtrup2020}.

\begin{table}[t]
\caption{Calculated representatives for injection molding processes of cover and plate part depending on process state.}
\begin{tabular}{p{0.75cm}p{0.85cm}p{1.075cm}p{0.9cm}p{1.2cm}p{1cm}p{0.85cm}p{1.075cm}p{0.9cm}p{1.2cm}p{1cm}}
\toprule
                   & \multicolumn{5}{l}{cover}                                      & \multicolumn{5}{l}{plate}                                      \\ \hline
Rep. No. & start-up & stable process & down-times & regrind material & DOE & start-up & stable process & down-times & regrind material & DOE \\ \hline
1                  & 569      & 432            & 342       & 122              & 739 & 854      & 536            & 308       & 479              & 726 \\
2                  & 76       & 95             & 706       & 946              & 353 & 0        & 544            & 200       & 807              & 620 \\
3                  & 491      & 719            & 972       & 683              & 421 & 223      & 386            & 269       & 183              & 181 \\
4                  & 9        & 906            & 600       & 451              & 33  & 459      & 628            & 886       & 393              & 457 \\
5                  & 0        & 544            & 701       & 588              & 640 & 149      & 145            & 342       & 38               & 245 \\
\bottomrule
\end{tabular}
\label{tab:case_study_representatives}
\end{table}

\subsubsection{Summaries}

Data summarizations are new means to generate added value for IMM operators. In industrial injection molding, operators are usually supervising not one, but multiple production machines. Consequently, when switching to a particular IMM, the operator will appreciate a short and comprehensive summary of what the injection molding cycles since his last visit do look like. Since, depending on cycle time, this may include hundreds or more cycles, manual analyses quickly become infeasible and computational algorithms are first choice.

\begin{figure}[t]
\centering
\resizebox{\textwidth}{!}{
\begin{tikzpicture}
	\begin{axis}[xlabel=Time $\left\lbrack s \right\rbrack$,ylabel=Melt pressure $\left\lbrack \text{bar} \right\rbrack$, width=\textwidth, y post scale=0.35, xmin=-0.5, xmax=15, ymin=-50, ymax=750]
		\addplot[color=niceblue, mark=none, line width=0.8pt] table[x=seconds, y=Inj1PrsAct, col sep=comma] {results/casestudy/38.csv};
		\addlegendentry{Cycle No. 38}

		\addplot[color=nicered, mark=none, line width=0.8pt] table[x=seconds, y=Inj1PrsAct, col sep=comma] {results/casestudy/183.csv};
		\addlegendentry{Cycle No. 183}

		\addplot[color=nicegreen, mark=none, line width=0.8pt] table[x=seconds, y=Inj1PrsAct, col sep=comma] {results/casestudy/393.csv};
		\addlegendentry{Cycle No. 393}

		\addplot[color=niceyellow, mark=none, line width=0.8pt] table[x=seconds, y=Inj1PrsAct, col sep=comma] {results/casestudy/479.csv};
		\addlegendentry{Cycle No. 479}

		\addplot[color=nicepurple, mark=none, line width=0.8pt] table[x=seconds, y=Inj1PrsAct, col sep=comma] {results/casestudy/807.csv};
		\addlegendentry{Cycle No. 807}

	\end{axis}
\end{tikzpicture}
}
\caption{Five representatives for plate part with varying regrind amount. Both maximum melt pressure during injection as well as plasticization time are affected.}
\label{fig:casestudy_representatives_plot}
\end{figure}
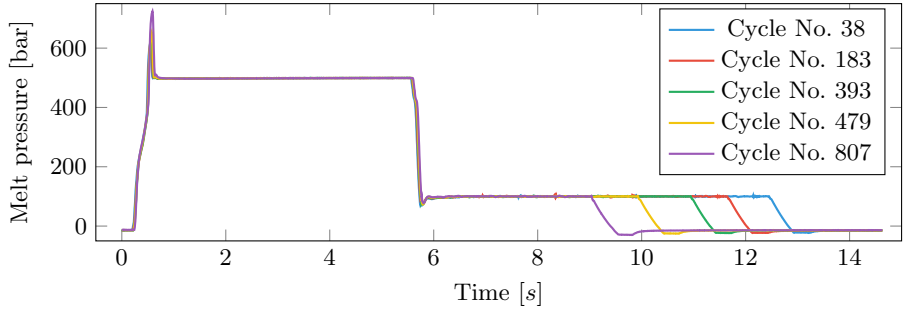

Today’s IMM controls record a variety of different sensor signals. One of the most important ones is the melt pressure since it contains detailed information on the melt viscosity. This holds true primarily in the injection phase, which is usually based on a controlled injection speed so that the melt pressure is a resulting parameter. However, the melt pressure timeseries may also provide information on the holding phase, decompression from holding to plasticization phase (decompression 1), as well as decompression after plasticization (decompression 2). Due to the high informative value, we chose the melt pressure for our analysis. Figure \ref{fig:casestudy_representatives_plot} shows how varying regrind amounts affect the melt pressure curve.

Since all important information can be found from injection phase until decompression 2, we sequenced all timeseries with the corresponding trigger signals. In the following, we evaluate how valuable data summarization of injection molding process data are. Therefore, we analyze to which extent the algorithmically found representatives match the expectations based on technical expertise regarding the induced process conditions in our datasets. Table \ref{tab:case_study_representatives} shows the first five representatives for each of them.

For the start-up process, the calculated representatives match the expectations based on process knowledge quite well: For both parts, the first representative is in the second half of the dataset. At this time, the process is already rather stable and the changes yielding to thermal equilibrium approach zero and a lot of cycles are similar to the chosen one. Furthermore, in both cases, the first cycle is among the top five representatives. This is also expected since at the very start of the experiment the differences from cycle to cycle are most significant. For the remaining chosen cycles, no clear system can be derived, nor would it be expected.

In the stable process, for both parts the found representatives are randomly distributed over the complete dataset. This also matches our expectation, as there are no systematic influences on the process, so none should show up in the data. As a matter of fact, if there was a system in the distribution of the representatives, this was a hint that there is a flaw in the experiment. 

In the downtime processes, some chosen representatives are directly after the downtimes and some in the middle of the cycles between two downtimes. This matches our expectations, since in general there should be majorly these two process states. Also important is the fact that the first chosen representative for both plastic parts is not directly after a downtime. This is due an asymptotic behavior taking place after restart, eventually yielding thermal equilibrium. Therefore, there should be many more cycles similar to some representative in amid than exactly at restart. 

Since in the process with regrind material, the regrind fraction is adapted every 200 cycles, there are five main process states expected which should manifest in one representative for each section. In practice, for both parts there are four different sections represented among the top five representatives, which is still a good result.

For the DOE, in total 43 operation points exist, with 20 injection molding cycles produced each point. Consequently, one would expect that each of the first 43 representatives would match the 43 operation points. Although this holds true for the first five representatives shown in the table, in total, for the cover, 33 sections are represented as well as 28 sections for the plate. The deviation can be explained by the fact that different factors in the DOE have the opposite effect, e.g., a high melt temperature lowers the viscosity and will thus yield a low pressure, while a high injection speed increases the pressure. If the two factors interact, the net effect might be rather small and variances within the operation point sections may be prevail, so more than one representative may be found in a specific section.

All in all, the calculated representatives agree very well with the expectations based on process knowledge. Considering that we only used the melt pressure timeseries and ignoring further sensor signals, the results a quite strong. Based on this, providing the machine operator with such data summarization may create added value. This could be further enhanced, if information on the root cause for the found representatives can be provided.

\section{Conclusion}
\label{sec:conclusion}

In this paper, we presented a novel GPU algorithm to evaluate the submodular function of Exemplar-based clustering.
Although exposing beneficial properties, its complexity could become prohibitive regarding the wall-clock run-time.

We briefly discussed the current state of practical applications of submodular functions and GPU accelerations of clustering. We formally established submodular functions and introduced the Greedy optimizer with its approximation guarantees. We then established Exemplar clustering and the accompanying submodular function by discussing $k$-medoids loss.

We claim that real-time applications demand results on time behavior that is measured in physical entities like seconds. The worst-case complexity determines the influential parameters. Systematic experiments result in wall-clock time for execution and allow to investigate the speedup of one implementation over another in varied settings of the parameters.  

We introduced our novel GPU algorithm, for which we engineered a work matrix first and discussed, how hardware features like shared memory might be used here. We especially took care of choosing an appropriate memory layout. Experiments have shown, that our algorithm succeeds in remarkable speedups of up to 72x using a workstation-grade hardware, while assuming FP32 computations and comparing to an adequate single-thread CPU implementation. Given, that a multi-threaded CPU algorithm represents the baseline, speedups ranging from 3.3x to 5.1x were possible. Moreover, we also reviewed the usage of FP16 arithmetic and revealed great speedups ranging from 16x to 415x compared to the FP32 CPU implementation, depending on whether single- or multi-thread computation was considered and which run-time-critical property was subject to variation. Furthermore, we investigated the use of embedded GPUs like the NVIDIA Jetson TX2 and were able to achieve speedups of up to 35.5x depending on the concrete problem to solve and floating point precision required. Overall, the GPU implementation results in considerable speedups.

We provided a comprehensive case study, which proves the usefulness of EBC-based summaries within the industrial domain of injection molding machinery. It was especially visible, that our summaries provided a diverse and extensive view on the underlying IMM process, which may helps operators with process control in modern plants with large arrays of machinery.

For future work it may be interesting to see, which dimensionality reduction techniques are appropriate for industrial process control, to reduce optimization times and to provide summaries even faster. It may be also interesting to see, which automatic decisions can be made using summaries within the IMM domain, to alert operators earlier of machine dysfunction.

\section*{Acknowledgments}

Part of the work on this paper has been supported by the German Competence Center for Machine Learning Rhine Ruhr (ML2R, 01IS18038A), funded by the German Federal Ministry for Education and Research and by Deutsche Forschungsgemeinschaft (DFG) within the Collaborative Research Center SFB 876 "Providing Information by Resource-Constrained Analysis", project A1.

\bibliographystyle{splncs04}
\bibliography{paper}

\end{document}

%% file: results/speedup_agg.tex
\begin{tabular}{llrrrrrrrrrrrr}
\toprule
           &     & \multicolumn{4}{l}{$N$} & \multicolumn{4}{l}{$l$} & \multicolumn{4}{l}{$k$} \\
           &     & \multicolumn{2}{l}{FP16} & \multicolumn{2}{l}{FP32} & \multicolumn{2}{l}{FP16} & \multicolumn{2}{l}{FP32} & \multicolumn{2}{l}{FP16} & \multicolumn{2}{l}{FP32} \\
           &     &     ST &    MT &    ST &   MT &     ST &    MT &    ST &   MT &     ST &    MT &    ST &   MT \\
\midrule
\multirow{3}{*}{Quadro vs. Xeon} & min &    8.5 &   0.8 &  34.0 &  3.3 &  273.9 &  20.3 &  68.3 &  4.8 &   61.2 &   4.3 &  47.1 &  3.3 \\
           & mean &  391.3 &  27.7 &  67.4 &  4.8 &  400.4 &  28.3 &  70.4 &  5.0 &  249.1 &  17.6 &  53.1 &  3.8 \\
           & max &  436.0 &  30.5 &  71.5 &  5.0 &  438.2 &  30.8 &  71.9 &  5.1 &  424.1 &  29.9 &  71.0 &  5.0 \\
\cline{1-14}
\multirow{3}{*}{TX2 vs. A72} & min &    5.1 &   1.3 &   4.3 &  1.5 &   24.3 &   6.2 &   5.7 &  1.5 &   26.6 &  12.3 &   4.7 &  2.2 \\
           & mean &   28.0 &  11.2 &   4.9 &  1.9 &   32.2 &  10.9 &   5.9 &  2.0 &   30.0 &  13.4 &   5.4 &  2.4 \\
           & max &   35.5 &  15.8 &   6.0 &  2.3 &   34.9 &  12.9 &   6.0 &  2.3 &   34.5 &  14.3 &   6.0 &  2.7 \\
\bottomrule
\end{tabular}